\documentclass[prc,aps,nofootinbib,tightenlines,superscriptaddress,floatfix,twocolumn,preprintnumbers]{revtex4}
\usepackage{bbm}
\usepackage{graphicx}
\usepackage{amsmath}
\usepackage{amsfonts,amsbsy}
\usepackage{amssymb}
\usepackage{subfigure}
\usepackage[usenames,dvipsnames]{color}
\newcommand{\gc}{{\gamma-\mathrm{ch}}}

\newcommand{\la}{\left<}
\newcommand{\ra}{\right>}
\newcommand{\lf}{\left(}
\newcommand{\rf}{\right)}
\newcommand{\lt}{\left[}
\newcommand{\rt}{\right]}
\newcommand{\pdf}{{\cal P}(f)}
\newcommand{\eff}{\varepsilon}
\newcommand{\ph}{\gamma}
\newcommand{\ch}{\mathrm{ch}}
\newcommand{\cont}{\frac{\varepsilon_{\ch,\ph}}{\varepsilon_{\ph}}}

\newcommand{\ngap}{\!\!\!\!\!}

\newcommand{\na}{\la N \ra}
\newcommand{\nc}{\la N_\ch \ra}
\newcommand{\np}{\la N_\ph \ra}
\newcommand{\ncp}{\la N_\ch N_\ph \ra}
\newcommand{\npf}{\la N_\ph(N_\ph-1) \ra}
\newcommand{\ncf}{\la N_\ch(N_\ch-1) \ra}

\newcommand{\nopm}{\la N_{\pi^0} N_{\pi^\pm} \ra}
\newcommand{\npmf}{\la N_{\pi^\pm}(N_{\pi^\pm}-1) \ra}
\newcommand{\nof}{\la N_{\pi^0} (N_{\pi^0} -1) \ra}
\newcommand{\npm}{\la N_{\pi^+}N_{\pi^-} \ra}

\newcommand{\be}{\begin{equation}}
\newcommand{\ee}{\end{equation}}
\newcommand{\bea}{\begin{eqnarray}}
\newcommand{\eea}{\end{eqnarray}}
\newcommand{\ndyn}{\nu_{dyn}}

\begin{document}
\title{Study of $\gamma$-charge correlation in heavy ion collisions, various approaches.}
\author{Prithwish Tribedy}
\author{Subhasis Chattopadhyay}
\affiliation{Variable Energy Cyclotron Centre, 1/AF Bidhan Nagar, Kolkata-700064, India}
\author{Aihong Tang}
\affiliation{Physics Dept., Brookhaven National Laboratory, Upton, NY 11973, USA}

\begin{abstract}
  Event-by-event $\gc$ correlation is used in studying systems going through QCD chiral phase transition. In this paper various methods for measuring $\gc$ correlation in heavy ion collisions have been discussed. 
  Dynamical fluctuation due to formation of domains of DCC that can affect $\gc$ correlation has been discussed. 
  We study known detector and statistical effects involved in these measurements 
  and suggest suitable robust observables $\Delta\ndyn$ and $r_{m,1}$ sensitive to small $\gc$ correlation signal. These variables are constructed based on moments of multiplicity distributions of photon and charged particles. Estimations of expected measurable signals of $\gc$ correlation from various available models such as for ideal Boltzmann gas of pions, monte-carlo models based on transport and mini-jets have been discussed. 
  Collision centrality dependence of the observables have been estimated from Central Limit Theorem and found to be consistent with the model predictions. 
  We find that observables show high sensitivity to fraction of DCC events and have nonlinear dependence on fraction of pions carrying DCC signals. Variation of $r_{m,1}$ with orders of its higher moments $m$ is a observable to extract the nature and strength of $\gc$ correlation.
  
\end{abstract}

\preprint{ }

\maketitle

\section{Introduction}
Based on decades of experimental searches and theoretical studies it is widely believed that high energy heavy ion collisions produce realistic scenario for studying the phase transition from hadronic matter to Quark gluon plasma. It is believed to be associated with two different transitions, de-confinement and restoration of QCD chiral symmetry. Fluctuation of conserved quantities has been proposed \cite{Jeon:1999gr} to be an important experimental signature for such phase transition.
%
Hadronic system that is mostly dominated by pions is expected to show a global isospin conservation. In such scenario the event-by-event isospin number fluctuation is an interesting observable. The QCD chiral phase transition is associated with melting of 4-vector 
condensates. An interesting phenomena like formation of metastable domains of ``Disoriented Chiral Condensate'' (DCC)  is predicted to occur due to the orientation of this condensate relative to the direction of its scalar component. Such a phenomena is possible for a scenario of rapid cooling like quenching~\cite{Bjd, Blaizot:1992at, Rajagopal:1992qz, Rajagopal:1995bc} for system going from chiral symmetry restored phase to broken phase.
  Formation of DCC domains causes anomalous production of charged or neutral pions depending on the orientation of vacuum towards its pseudo scalar component. Such phenomena might survive final state interactions and appear in the form of multiplicity fluctuation of pions of relative isospins\cite{Rajagopal:1992qz}. As the detected charged and neutral particles are mostly from the charged pions and the decay of neutral pions respectively this would appear in the form of $\gc$ anti-correlation. 
 The experimental searches of DCC so far includes searches in pp collisions\cite{Brooks:1999xy}, cosmic ray events\cite{Lattes:1980wk} and in heavy ion collisions\cite{Aggarwal:1997hd, Appelshauser:1999ft}. Our discussion would be relevant to the search in heavy-ion collisions. The prediction 
for a hot medium described by the linear sigma model~\cite{Krzywicki:1998sc} showed that in the case of central collision of Pb-Pb at SPS energies, the likelihood of the DCC events is less than $10^{-3}$. Experimental searches at SPS WA98 experiment \cite{Aggarwal:1997hd,
Aggarwal:2000aw, Aggarwal:2002tf, Collaboration:2011rsa} at $\sqrt{s}$=17.3 GeV estimated an upper limit of $3\times10^{-3}$. It has been argued \cite{Rajagopal:2000yt}  that in case of rapid cooling like quenching scenario, higher collision energies corresponding to lower chemical potential (e.g. $\mu_{RHIC} < \mu_{SPS}$) provides faster cooling rate ($\left| dT/dt \right|$). This suggests that RHIC and LHC collisions provide more favorable condition for DCC production than SPS collisions. 

 From experimental point of view such a study is associated with simultaneous measurement of photons and charged particles in common phase space with very high sensitivity at low momentum. This is because the decay of domains of DCC are final stage phenomena of the evolution of heavy ion collision and the pions carrying signals are expected to be of low momentum. A combination of pre-shower Photon Multiplicity detector(PMD)\cite{STARNIM:PMD} and forward time projection chamber(FTPC)\cite{STARNIM:FTPC} at STAR experiment at RHIC and Photon Multiplicity detector(PMD) and Forward Multiplicity Detector(FMD) at ALICE experiment\cite{Aamodt:2008zz} at LHC have the required criteria to satisfy such goal. 
 
 In this paper we would like to highlight few issues associated to $\gc$ correlation analysis and propose a method. We use generating function approach to calculate different variables and include various detection effects like efficiencies, effect of mis-identification etc. Observables of $\gc$ correlation are constructed to be suitable for heavy ion collisions that can disentangle dynamical fluctuation. Assuming formation of DCC domains to be one of the probable sources of dynamical signal of isospin fluctuation we discuss the sensitivity of the observables to the fraction of DCC events and the fraction of DCC candidates in an event. 
 Relevant to the heavy ion collisions we discuss the centrality dependance of the variables.
 We estimate $\gc$ correlation from various models and implement a DCC-model based on HIJING event generator.
 
 In the section \ref{sec_method} we outline the method of construction of the observables and their values for DCC events of varying fraction. Section \ref{sec_misid}, \ref{sec_reso} and \ref{sec_clt} describe the detector effect like mis-identification, the role of resonances and centrality dependance respectively on the proposed variables. In section \ref{sec_pifrac} we have calculated the sensitivity of the variables on DCC event fraction and pion fractions in DCC events. For studying the experimental sensitivity of DCC, we have studied various non-DCC models in section \ref{sec_model} and implemented DCC in a Monte-Carlo based events in section \ref{sec_dccmodel}. We summarize in section \ref{sec_sum}. 
\section{Method}
\label{sec_method}
Fluctuation of particle ratios has been addressed previously in case of conserved quantities like net strangeness in terms of kaon-to-pion ratio and net baryons in terms of proton to pion ratios. Relevant to our case is the study of photon to charge particle multiplicity ratio. Observables used in such cases are designed in such a way so as to eliminate the statistical fluctuations and at the same time be robust against detector inefficiency. A simple way of implementing detector efficiencies in terms of a binomial probability distribution function say of the form $P(n,N,\eff)= \!\!^N\mathrm{C}_n\,\eff^n (1-\eff)^{N-n}$ would reveal the fact that the second moment of observed multiplicity $n$ is not proportional to second moment of produced multiplicity $N$. The efficiency term $\varepsilon$ does not factorize for quantities like variance, skewness and kurtosis\footnote{variable $D= 4\la\Delta Q^2\ra/N_\ch$ where $Q^2$ is the variance of the net charge($N_+-N_-$) in Ref\cite{Jeon:1999gr} gives different values for QGP and pion gas but depends on efficiency.}. However the quantities like observed second and higher order factorial moments comes out to be proportional to the measured corresponding factorial moments like $\langle n(n-1)\rangle =\eff^2\langle N(N-1)\rangle$.
 Ratios of various factorial moments with powers of mean multiplicity would simply cancel the explicit efficiency dependence. In case of correlation of multiplicities, there could be more complicated detector effects like mis-identification of one species in the form of another, decay and resonance production. This could lead to spurious correlation affecting the final results.
Also in case of heavy ion collisions there are centrality and system size dependence. If heavy-ion collisions are assumed to be linear superpositions of multiple hadronic collisions, then variables are supposed to show number of source scaling~\cite{Luo:2010by}. Based on similar context and considering various other aspects of particle ratio-fluctuation, two observables were introduced earlier as measure of $\gc$-ratio fluctuations. $\nu_{dyn}^\gc$ was introduced in Ref~\cite{nudyn} and used by STAR Collaboration~\cite{star_kpi,star_dcc} and $r_{m,1}^{\gc}$ was introduced by Minimax collaboration\cite{Minimax}. The variable $\ndyn$ is defined as 
\be
\nu_{dyn}^\gc=\frac{\ncf}{\nc^2}\,+\,\frac{\npf}{\np^2}\,-\,2\frac{\ncp}{\np\nc}
\label{eq_ndyn}
\ee 
 which for Poissonian case should give zero. The variable $r_{m,1}$ is defined as 
\be
 r_{m,1}^\gc=\frac{\la N_\ch(N_\ch-1)..\lf N_\ch-m+1\rf \, N_\ph \ra \nc}{\la N_\ch(N_\ch-1)..(N_\ch-m)\ra \np}.
\label{eq_rm1}
\ee
It is designed such that for all the moments it gives a value equal to 1 for Poisson case and higher order moments show larger sensitivity to signals. In this section we would like to discuss the applicability, robustness and sensitivity of these two variable for $\gc$ correlation. 
Since we are interested in fluctuation of ratio of multiplicities let us consider  $f=N_{\pi^0}/(N_{\pi^0}+N_{\pi^\pm})$ to be the neutral pion fraction. The idea is that by using proper combinations of moments we can eliminate the efficiency dependence and express our observable in terms of the fluctuation of the fraction $f$.
 The most efficient way of studying the moments including the dynamical and detector effect is to follow the generating function approach~\cite{Minimax} where we define,
\begin{equation}
G(z)=\sum\limits_{N=0}^{\infty}z^N\,P(N) 
\end{equation}
where $P(N)$ denotes the distribution of parent multiplicity where , $N=N_0+N_{ch}$ denotes sum of all neutral and charged pions. Different moments are calculated by taking derivatives of $G(z)$ w.r.to $z$ evaluated at $z=1$.
%
 Considering the fact that the neutral pions are distributed according to the probability $\pdf$ the generating function has to be modified accordingly 
\begin{equation}
G(z_{\ch},z_0) = \int\limits_0^1 \, df \, {\cal P}(f) \sum \limits_N P(N) \lt fz_0\,+\,(1-f)z_{\ch}\rt^N.
\label{eq_genf}
\end{equation}
 The distribution $\pdf$ is the event-by-event measured distribution of neutral pion fraction. Isospin symmetry for a pion gas corresponds to a generic case of pion productions for which $\pdf= \delta(f-1/3)$. In case of DCC like events\cite{Blaizot:1992at,Rajagopal:1995bc}  we have $\pdf=1/2\sqrt{f}$. For propagation of generating function to include the decay of neutral pions to observed photons we apply the ``cluster decay theorem''~\cite{Pumplin}. We can express the overall generating function as
\begin{equation}
G_{obs}\lf z_{\ch},\, z_{\gamma}\rf =G \lf g_{\ch}\lf z_{\ch}\rf ,\, g_0\lf z_{\gamma}\rf \rf
\label{eq_gobs}
\end{equation}
where $g_0(z_\gamma)=z_{\gamma}^2$ considering the fact that every neutral cluster decays into two photons and the charge particles do not decay, $g_{\ch}(z_{\ch})=z_{\ch}$. To make the scenario more realistic and taking the advantage of same  theorem, one can include detection efficiencies into the final form of generating function. We consider the observing and non-observing as different decay modes with probability equal to the detection efficiency. So for charged and neutral clusters we redefine
\begin{eqnarray}
\label{eq_gchg0}
g_{\ch}(z_{\ch})&=&(1-\eff_{\ch})+\eff_{\ch}z_{\ch} \\
g_{0}(z_{\gamma})&=& ((1-\eff_\ph) \,+\, \eff_\ph z_\ph)^2 \nonumber
\end{eqnarray}
 Here $\eff_{\ch}$ is the efficiency of charge particle detection and $\eff_{\gamma}$ is the efficiency of detecting a photon coming from decay of a neutral pion. We can calculate various factorial moments of multiplicity with detector efficiency folded in terms of derivatives of final generating function. We can define a generalized factorial moment as 
\begin{equation}
\mathrm{f}_{m,n}\!\!=\left. \frac{\partial^{m,n} G_{\mathrm{obs}}(z_{\ch},z_{\ph})}{\partial z_{\ch}^m \, \partial z_{\ph}^n }\right|_{z_{\ch}=z_{\ph}=1} \ngap\ngap= \la\! \frac{N_{\ch}! \, \, N_{\ph}! }{(N_{\ch}-m)! \, (N_\ph-n)!}\!\ra
\label{eq_gnmom}
\end{equation}
 It is convenient to express our observables given in eq.\ref{eq_ndyn} and eq.\ref{eq_rm1} in terms $\mathrm{f}_{m,n}$ as
\be
\nu_{dyn}^\gc=\frac{\mathrm{f}_{20}}{\mathrm{f}_{10}^2}\,+\,\frac{\mathrm{f}_{02}}{\mathrm{f}_{01}^2}\,-\,2\frac{\mathrm{f}_{11}}{\mathrm{f}_{10} \, \mathrm{f}_{01}}\,\,\,\,,\,\,\,\, r_{m,1}^\gc =\frac{\mathrm{f}_{m1}\, \mathrm{f}_{10}}{\mathrm{f}_{(m+1)0} \, \mathrm{f}_{01}}
\label{eq_variables_fmn}
\ee
 Using eq.~\ref{eq_genf}, eq.~\ref{eq_gobs} and eq.~\ref{eq_gnmom} we can express few factorial moments in terms efficiency and average of neutral pion fraction.
 \bea
\mathrm{f}_{10}&=&\la 1-f \ra \eff_\ch \na \nonumber \\
%
\mathrm{f}_{01}&=&\la f \ra 2\eff_{\ph} \la N \ra \nonumber \\
%
\mathrm{f}_{11}&=&\la f \lf 1-f \rf \ra 2\eff_{\ph} \, \eff_\ch \la N\lf N-1\rf \ra \nonumber \\
%
\mathrm{f}_{20}&=&\la \lf 1-f \rf ^2\ra\eff_\ch^2 \la N \lf N-1 \rf \ra \nonumber \\
%
\mathrm{f}_{02}&=&\la f^2\ra 4 \eff_{\ph}^2 \la N \lf N-1\rf \ra +2\eff_{\ph}^2 \la f \ra \la N \ra \nonumber
\eea
 Substituting these in eq.\ref{eq_ndyn} we obtain
\bea
\nu_{dyn}^{\ph-\ch}=&\lf \frac{\la(1-f)^2\ra}{\la 1-f \ra^2}\,+\, \frac{\la f^2\ra}{\la f \ra^2} \,-\,2\,\frac{\la f(1-f)\ra}{\la f \ra \la 1-f \ra} \rf \frac{\la N(N-1)\ra}{\la N \ra^2} \nonumber \\ 
& \,+\, \frac{1}{2\la f \ra \la N \ra}.
\label{eq_ndyneff}
\eea
  We note here that the for generic case the term inside the bracket is zero and we have 
 \be
\left.\ndyn^{\ph-\ch}\right|_{generic}= \frac{1}{2\la f \ra \la N \ra}.
\ee
Using proper combination of factorial moments and doing a simple method of event mixing one can extract the generic value of $\ndyn^{\ph-\ch}$ (see appendix-\ref{app_mixevnt} for details).
Subtracting the generic value of $\ndyn^{\ph-\ch}$ one can get rid of the last term in eq.\ref{eq_ndyneff}. 
 

 %
 So we propose a modified variable $\ndyn-\ndyn^{generic}$ given by
\bea
\Delta\ndyn^{\ph-\ch}= &&\lf \frac{\la(1-f)^2\ra}{\la 1-f \ra^2}\,+\,\frac{\la f^2\ra}{\la f \ra^2} \,-\,2\,\frac{\la f(1-f)\ra}{\la f \ra \la 1-f \ra} \rf  \nonumber \\
&&\times \frac{\la N(N-1)\ra}{\la N \ra^2}.
\label{eq_delndyn}
\eea
 
In ideal scenarios when all the particles are detected one can approximate $g_0(z_\gamma)=z_{\gamma}^2$ and $g_{\ch}(z_{\ch})=z_{\ch}$. In that case one can show using eq.\ref{eq_gobs} and eq.\ref{eq_gnmom} that
\be
\left.\ndyn^{\ph-\ch}\right |_{generic}=\frac{1}{2\la N \ra \la f \ra}\approx\frac{1}{\sqrt{\la N_\ch \ra \la N_\ph \ra}}
\ee 
 irrespective of any value of $\ndyn$. So in that case the observable $\Delta\ndyn$ can be estimated to be 
 \be
 \Delta\ndyn^{\ph-\ch}=\ndyn^{\ph-\ch} \,-\, \frac{1}{\sqrt{\la N_\ch \ra \la N_\ph \ra}}
 \label{eq_delndyn_simple}
 \ee

	Following the same approach one can express the variable $r_{m,1}$ as
\be
r_{m,1}^{\ph-\ch}=\frac{\la f(1-f)^m\ra \la 1-f\ra}{\la (1-f)^{m+1}\ra \la f \ra}.
\label{eq_rm1eff}
\ee 
 Now we would like to discuss the sensitivity of these two variables for a given fraction of DCC like signal. If $x$-fraction of events has DCC like domain formation, in simplistic case one can assume that the distribution of neutral pion fraction to be a combination of generic and DCC probability distribution given by
\be
{\cal P}(f) = x \, \frac{1}{2 \sqrt{f}} \,+\, (1-x) \, \delta\lf f-\frac{1}{3}\rf.
\label{eq_pdf}
\ee
 So for $\Delta\ndyn$ we have from eq.\ref{eq_delndyn}
\bea
\Delta\ndyn^{\ph-\ch}=&&\ngap\left. \lf \frac{\la(1-f)^2\ra}{\la 1-f \ra^2}\,+\,\frac{\la f^2\ra}{\la f \ra^2} \,-\,2\,\frac{\la f(1-f)\ra}{\la f \ra \la 1-f \ra} \rf \right|_{signal} \nonumber \\
&&\times \frac{\la N(N-1)\ra}{\la N \ra^2} \nonumber \\
=&&\frac{x}{5/9} \, \frac{\la N(N-1)\ra}{\la N \ra^2}
\label{eq_ndynsig}
\eea
 which is proportional to the fraction of DCC-like events. $\Delta\ndyn$ shows very high sensitivity to DCC like signal but it has dependency on the parent multiplicity and consequently to the collisions centrality. In later section we would discuss this issue in detail. In case parent distribution is Poissonian, the fluctuation term $\la N(N-1)\ra/\la N \ra^2$ would be equal to 1 giving $\Delta\ndyn^{\ph-\ch}\,\sim\,x/(5/9)$. 

  The robust observable expressed in eq.\ref{eq_rm1eff} would have a very particular $x$ dependence given by 
\be
r_{m,1}^{\ph-\ch}=1-\frac{m x}{(m+1)} F(m,x)
\label{eq_rm1sig}
\ee
where the function $F(m,x)$ is given by
\be
F(m,x)=\frac{1}{x\,+\,(1-x)\frac{2}{\sqrt{\pi}}\,\lf\frac{2}{3}\rf^{m+1}\,\frac{\Gamma(m+5/2)}{\Gamma(m+2)}}.
\label{eq_fmx}
\ee
 For ideal DCC case ($x$=1), the function $F(m,x)$=1 for all values of $m$. That gives $r_{m,1}=1/(m+1)$. For generic case($x=0$), $r_{m,1}$=1. Fig.1 shows the sensitivity of $r_{m,1}$ for small signals of DCC. 
 The functional form given in eq.\ref{eq_rm1sig} can be used to extract $x$ from a fit of $r_{m,1}$ with $m$. In the derivation of eq.\ref{eq_ndynsig} and eq.\ref{eq_rm1sig} we have assumed that the parent multiplicity distribution are similar for both the generic and DCC case and the efficiency factors are constant and independent of multiplicity and other kinematic parameters. 

\begin{figure*}
\begin{center}
\subfigure[Variation of $r_{m,1}$ with fraction of DCC signal]{
\includegraphics[height=7cm, width=7cm]{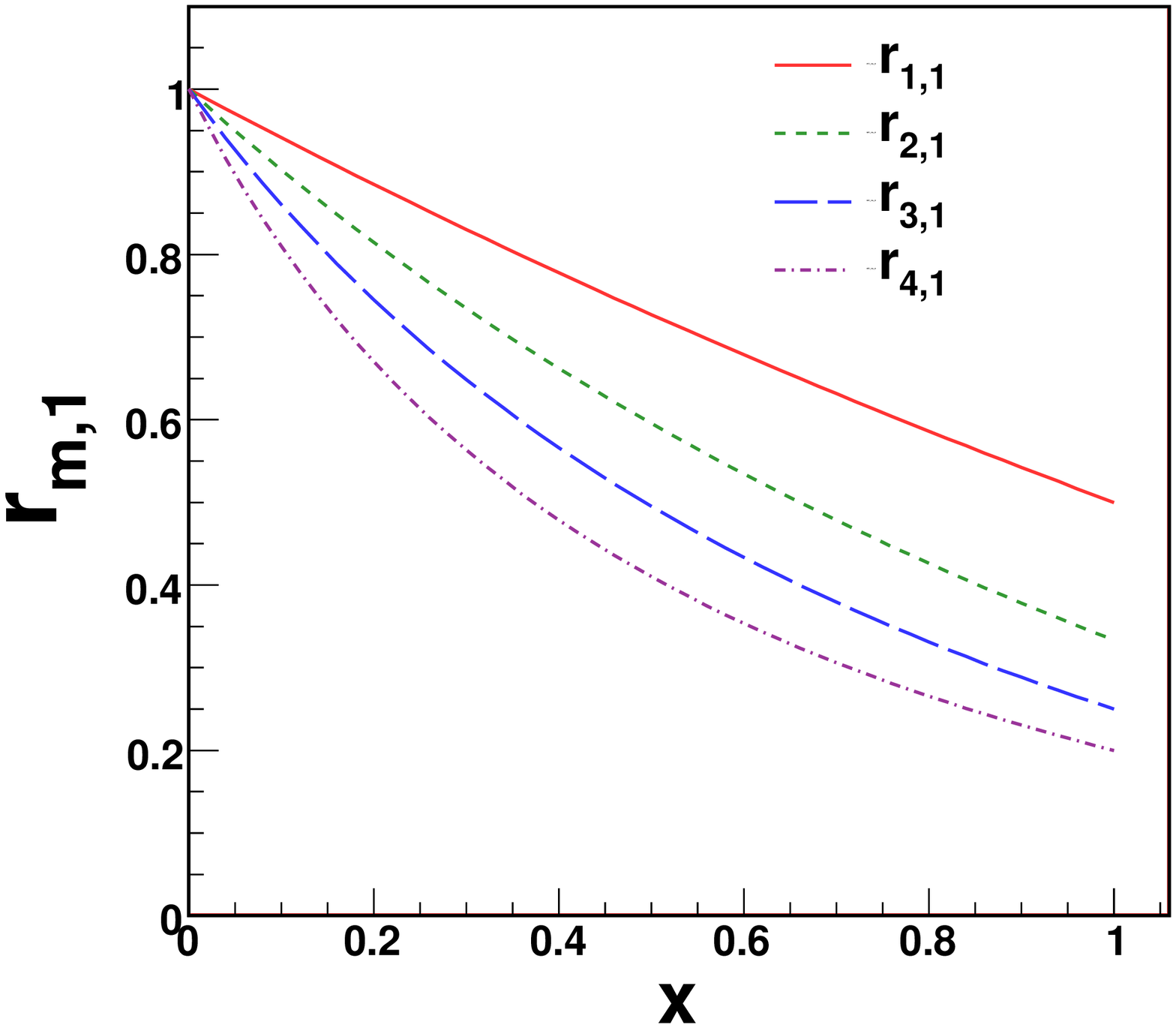}}
\subfigure[Variation of $r_{m,1}$ with higher order moments]{
\includegraphics[height=7cm, width=7cm]{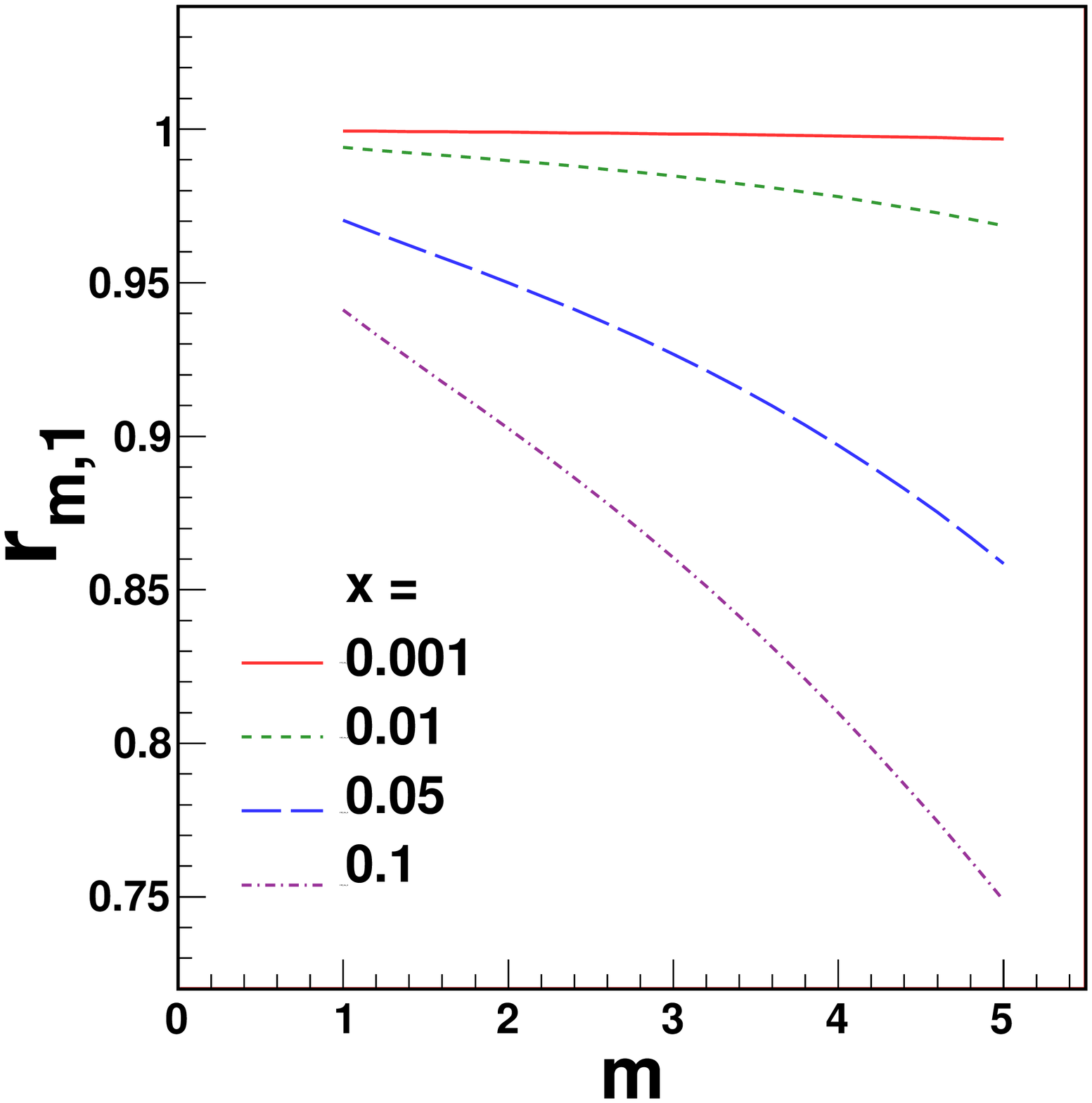}}
\caption{\label{fig_rm1} Sensitivity of variable $r_{m,1}$ and its higher moments. The higher order shows more sensitivity to small signals of anti-correlation.}
\end{center}
\end{figure*}
\section{Effect of mis-identification}
\label{sec_misid}
There are additional complications in realistic scenarios that have not been taken care of in the above prescriptions. The study of $\gc$ correlation is often complicated by mis-identification of charge particles by photons and vice versa. High energy depositions of charged hadrons can form a cluster in photon detector. Similarly photon conversion can show up as single or doubly detected tracks or clusters in charge particle detectors. In both the cases the observables are affected. Following the approach of the application of cluster decay theorem discussed in previous section, we obtain the modified forms of the generating functions 
\begin{eqnarray}
g_{\ch}(z_{\ch},z_\ph)&=&(1-\eff_{\ch}-\eff_{\ch,\ph})+\eff_{\ch}z_{\ch}+\eff_{\ch,\ph}z_\ph \nonumber \\
g_{0}(z_{\ch},z_{\ph})&=&\lf(1-\eff_{\ph}-\eff_{\ph,\ch}-\eff_{\ph,2\ch})+\eff_{\ph}z_{\ph} \right .\nonumber \\
&&+\left. \eff_{\ph,\ch}\,z_{\ch}+\eff_{\ph,2\ch}\,z_{\ch}^2\rf^2,
\end{eqnarray}
 where we view neutral pions decay with $100\%$ ``efficiency'' into two photons which themselves ``decay'' with a few modes. $\eff_\ch$ and $\eff_\ph$ are the efficiencies of detecting a charged particle and a photon, respectively. $\eff_{\ch,\ph}$ is the probability of charged particle being identified as a photon cluster and $\eff_{\ph,\ch},\, \eff_{\ph,2\ch}$ are the probability of a photon being identified as one and two charged particles, respectively. 
 Substituting these in eq.\ref{eq_gobs} one can calculate factorial moments folded with the contamination effect. 
The factorial moments are expressed as,
\bea
\label{eq_fmom_eff}
\mathrm{f}_{10}&=&\la(1-f)\eff_\ch \,+\, 2f \lf \eff_{\ph,\ch} \, + \, 2\eff_{\ph,2\ch} \rf \ra \na \nonumber \\
\mathrm{f}_{01}&=&\la (1-f)\eff_{\ch,\ph} \, + \, 2f\eff_\ph \ra \na \nonumber \\
\mathrm{f}_{11}&=&\la N(N-1) \, \lf (1-f)\eff_{\ch} \, + \, 2f \lf \eff_{\ph,\ch}+2\eff_{\ph,2\ch}\rf \rf \right. \nonumber \\
&& \times  \left.  \lf(1-f)\eff_{\ch,\ph}\,+\,2f\eff_\ph \rf + 2Nf \eff_\ph\lf \eff_{\ph,\ch} + 2 \eff_{\ph,2\ch} \rf \ra \nonumber \\
\mathrm{f}_{20}&=&\la N(N-1) \, \lf (1-f)\eff_\ch+2f(\eff_{\ph,\ch}+2\eff_{\ph,2\ch})\rf ^2 \, \right. \nonumber\\
&& \left. + \, 2Nf \, \lf 2\eff_{\ph,2\ch} +(\eff_{\ph,\ch}+2\eff_{\ph,2\ch})^2\rf \ra \nonumber \\
\mathrm{f}_{02}&=&\la N(N-1)\! \lf (1-f)\eff_{\ch,\ph}+2f\eff_\ph \rf ^2 \,+\, 2Nf \eff_\ph^2\ra
\eea
 This would lead to very complicated (see appendix-\ref{app_misid}) dependencies of $\Delta\ndyn$ and $r_{m,1}$ on various efficiency factors. However a relatively simple form can be obtained in the limit of small values of $\eff_{\ph,\ch}$ and $\eff_{\ph,2\ch}$. So in case of small photon conversion in the charged particle detector one can express $\Delta\ndyn$ as 
\begin{widetext}
\begin{equation}
\Delta\ndyn^{\ph-\ch}= \lf \frac{\la(1-f)^2\ra}{\la 1-f \ra^2} \,+\, \frac{ \la \lf(1-f)\cont+2f\rf^2 \ra}{\la(1-f)\cont+2f\ra^2} \,-\,2\, \frac{\la (1-f)\lf (1-f)\cont+2f \rf \ra}{\la1-f\ra \la (1-f)\cont+2f \ra} \rf \, \frac{\la N(N-1)\ra}{\la N \ra^2}
 \label{eq_delndyn_eff}
\end{equation}
\end{widetext}
%
where the the generic value of $\ndyn$ in this case will be given by
\be
\left. \ndyn^{\ph-\ch} \right |_{generic}\,=\, 
\frac{1}
{2\la f \ra \la N \ra \lf \cont+1\rf} 
\label{eq_ndyngen_cont}
\ee
which is small number for large values of $\la N \ra$. The robust variable $r_{m,1}$ can be represented as
\be
r_{m,1}^{\ph-\ch}=\frac{\la (1-f)^m \lf (1-f)\cont+2f \rf \ra \, \la 1-f \ra }{\la(1-f)^{m+1}\ra \,\la(1-f)\cont+2f\ra}.
\label{eq_rm1_eff}
\ee
Unlike previous case it is not possible to eliminate the efficiency factors in eq.\ref{eq_delndyn_eff} and eq.\ref{eq_rm1_eff}. So in this case if we want to analyze the sensitivity of those variables to $x$-fraction of DCC signals, we can use eq.\ref{eq_pdf} to obtain the modified forms as
\bea
\Delta\ndyn^{\ph-\ch}&=&\frac{x}{5/9}\,\frac{1}{\lf\cont+1\rf^2}\,\, \frac{\la N(N-1)\ra}{\la N \ra^2} \nonumber \\
r_{m,1}&=&1-\frac{mx}{m+1}\frac{1}{\lf \cont+1\rf}\,F(m,x).
\eea
 where $F(m,x)$ is given by eq.\ref{eq_fmx}. We can see that mis identification of charged particle as photon reduces the effective fraction of DCC events. The contamination factor appears as a ratio of $\eff_{\ch,\ph}/\eff_\ph$ keeping the functional form of the variables\,(eq.\ref{eq_ndynsig},\,eq.\ref{eq_rm1sig}) unchanged. We note here that $\Delta\ndyn^{\ph-\ch}$ has quadratic dependence on contamination factor whereas $r_{1,1}$ is affected only by a linear factor. This is because $\Delta\ndyn^{\ph-\ch}$ contains an extra photon fluctuation term absent in $r_{m,1}$.\\
\section{Resonance effect} 
\label{sec_reso}
{\color{Black} Resonance decays like $\rho \rightarrow \pi^\pm \ph $ is equivalent to artificial increase of pions and photons from generic case. Decays like $\omega \rightarrow \pi^0 + \pi^\pm$ would give rise to correlation in the pions. Overall effect of resonance would be equivalent to event-by-event fluctuation of charged or neutral particles. The effect of resonance leading to increase in photon and charged particle multiplicity could be considered to be equivalent to increase in efficiency of photon and charged particle detection. For event-by event fluctuations of efficiency would affect the observables, for e.g. the variable $r_{m,1}$ given in eq.\ref{eq_rm1eff} will be modified as
\be
r_{m,1}^{\ph-\ch}=\frac{\la f \lf 1-f \rf ^m\ra \la 1-f\ra}{\la  \lf 1-f \rf ^{m+1}\ra \la f \ra}  \frac{\la \eff_\ph  \eff_\ch^m \ra\la\eff_\ch\ra}{ \la\eff_\ch^{m+1}\ra \la \eff_\ph \ra}.
\ee
 It is difficult to conclude the behavior of the variables from the above expressions without putting a realistic number for the efficiencies. To study the effect of resonances in a more detailed way (sec.\ref{sec_model}, \ref{sec_dccmodel}) we have used Monte-Carlo models in which resonances are included.

\section{Centrality dependence}
\label{sec_clt}
In heavy ion collisions, signals are expected to have centrality dependence, it is therefore 
important to study 
the centrality dependence of the $\gc$ correlation. In a heavy ion collision, let us consider $N_{S}$ numbers of identical sources are responsible for particle production. If $N_{i}$ is the number of particles produced from $i$-th source, any variable $V(N_i)$ will have a distribution identical for all the sources. If we assume heavy-ion collision to be a linear superposition of many identical nucleon-nucleon collisions, under identical source approximation we can calculate the centrality dependence of the variable using ``central limit theorem''(CLT) \cite{Luo:2010by}. From CLT it follows that mean and variance of multiplicity would be given by
\bea
M(N)=M\lf \sum\limits_i^{N_S} N_i\rf = \sum\limits_i^{N_S} M(N_i)=N_S\,M({N}_i) \nonumber \\
\sigma^2(N)=\sigma^2\lf \sum \limits_i^{N_S} N_i\rf= \sum\limits_i^{N_S} \sigma^2(N_i) =N_S \, \sigma^2({N}_i).
\eea
 So from CLT we have the dependence $M(N)=\alpha N_S$ and $\sigma(N)=\beta \sqrt{N_S}$. Let us assume $N$ to be equal to the total number of produced pions where we have $N_\pi=aN_\ch\,+\,bN_\ph$. $N$ could also refer to individual number of photons or charged particles. In that case similar argument also holds for $M(N_\ch) \sim \alpha_1 N_S$ and $M(N_\ph) \sim \alpha_2 N_S$. The variance of total numbers of pions would give
\bea
\sigma^2(N_\pi) = \lf \la N_\pi^2 \ra -\la N_\pi \ra^2\rf \sim \beta^2 N_S \\
\la N_\pi^2\ra = \la (a N_\ch +b N_\ph)^2\ra \sim \beta_1 N_S +\beta_2 N_S^2
\eea
 and if we express pion multiplicity in terms of charged and photons we get,
\bea
\la N_\ch^2 \ra \propto \beta_1 N_S +\beta_2 N_S^2 \nonumber \\
\la N_\ph^2 \ra \propto \beta_1 N_S +\beta_2 N_S^2 \nonumber \\
\la N_\ch N_\ph \ra \propto \beta_1 N_S +\beta_2 N_S^2
\eea
 So from eq.\ref{eq_ndyn} and eq.\ref{eq_rm1} we can calculate the centrality dependence of the observables. For $\ndyn$ one has
\be
\ndyn^{\ph-\ch}\sim A\,+\,\frac{B}{N_S}\,  \equiv \, A^\prime \,+\,\frac{B^{\prime}}{\sqrt{\la N_\ph \ra \la N_{\ch} \ra}}
\ee 

 which is in fact the centrality dependence of all three terms in eq.\ref{eq_ndyn}. Here we note that  the constants $A^\prime$ and $B^\prime$ could be either positive or negative depending on which term in eq.\ref{eq_ndyn} is dominant. The variable $\Delta\ndyn$ would have the similar centrality dependence which is obvious from the form of eq.\ref{eq_delndyn_simple}. In heavy ion collisions, number of source participating in particle production can also be assumed to be proportional to number of participants ($N_S\sim N_{part}$) of the collision. In that case $\ndyn^{\ph-\ch}$ is expected to show a scaling behavior of the form $A+B/X$ with $X$ being either observed multiplicity or a Galuber variable $N_{part}$. In case of experimental measurements it is more convenient to express fluctuation variables in terms of measured multiplicities. 
  
 Based on similar approach one can comment on the centrality dependence of the robust observable. In the most general case one can have
\be
r_{m,1}=\frac{\sum\limits_p^m \alpha_p N_S^p}{\sum \limits_p^m \beta_p N_S^p}
\ee
 which shows identical dependence in both numerator and denominator. So according to CLT,  behavior of $r_{m,1}$ with multiplicity depends on the coefficients $\alpha_i$ and $\beta_i$. However it must be noted that breakdown of scaling from CLT would have several implications. 
The picture of identical source emission may not be valid in the case for formation of domains of DCC. In that case one might observe deviation from proposed scaling. 
  %
\begin{figure*}
\begin{center}
\subfigure[Variation of $r_{1,1}$ with multiplicity]{
\includegraphics[height=7cm, width=7cm]{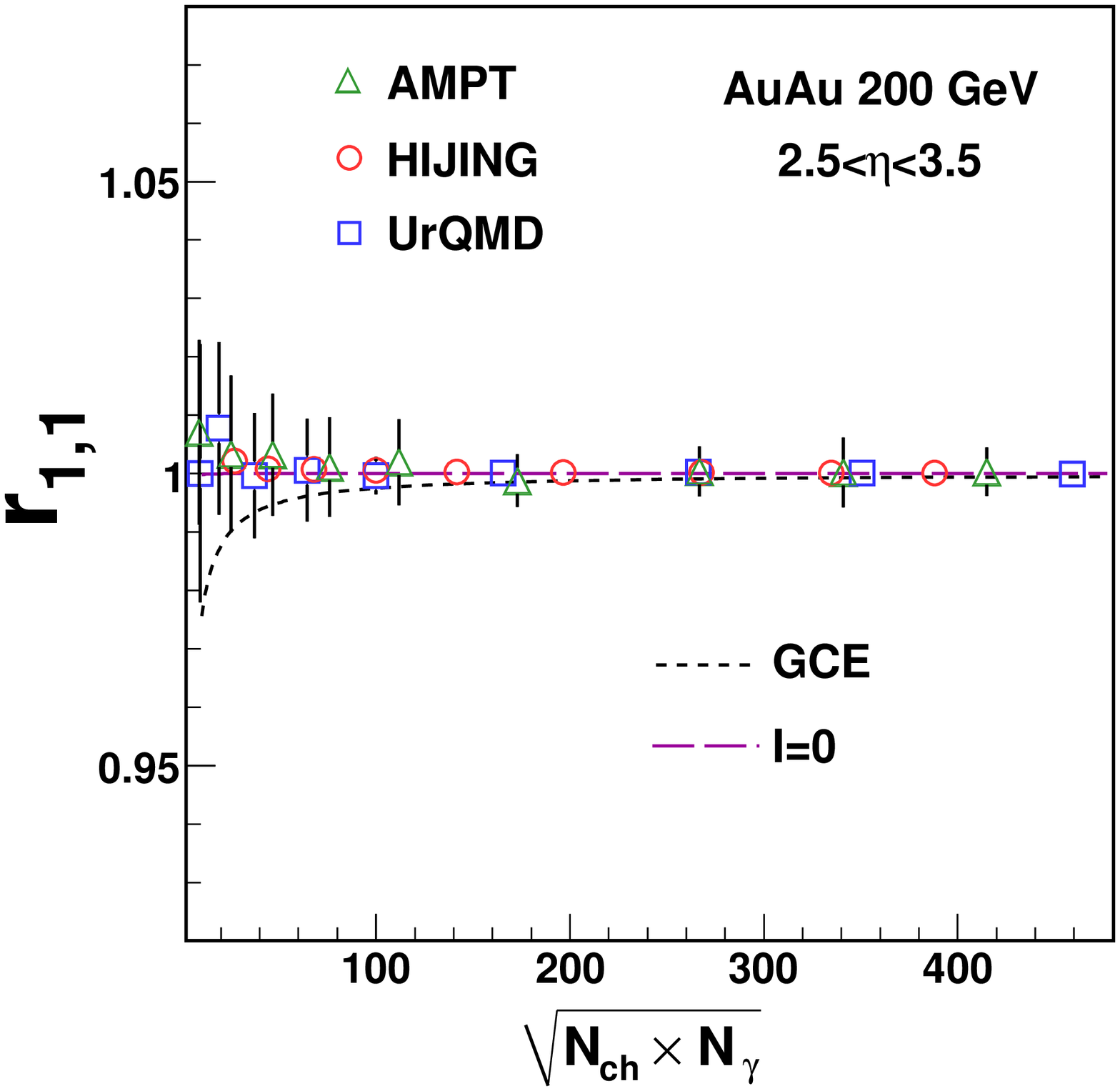}
\label{fig_r11_mult}}
\subfigure[Variation of $\Delta\ndyn$ with multiplicity]{
\includegraphics[height=7cm, width=7cm]{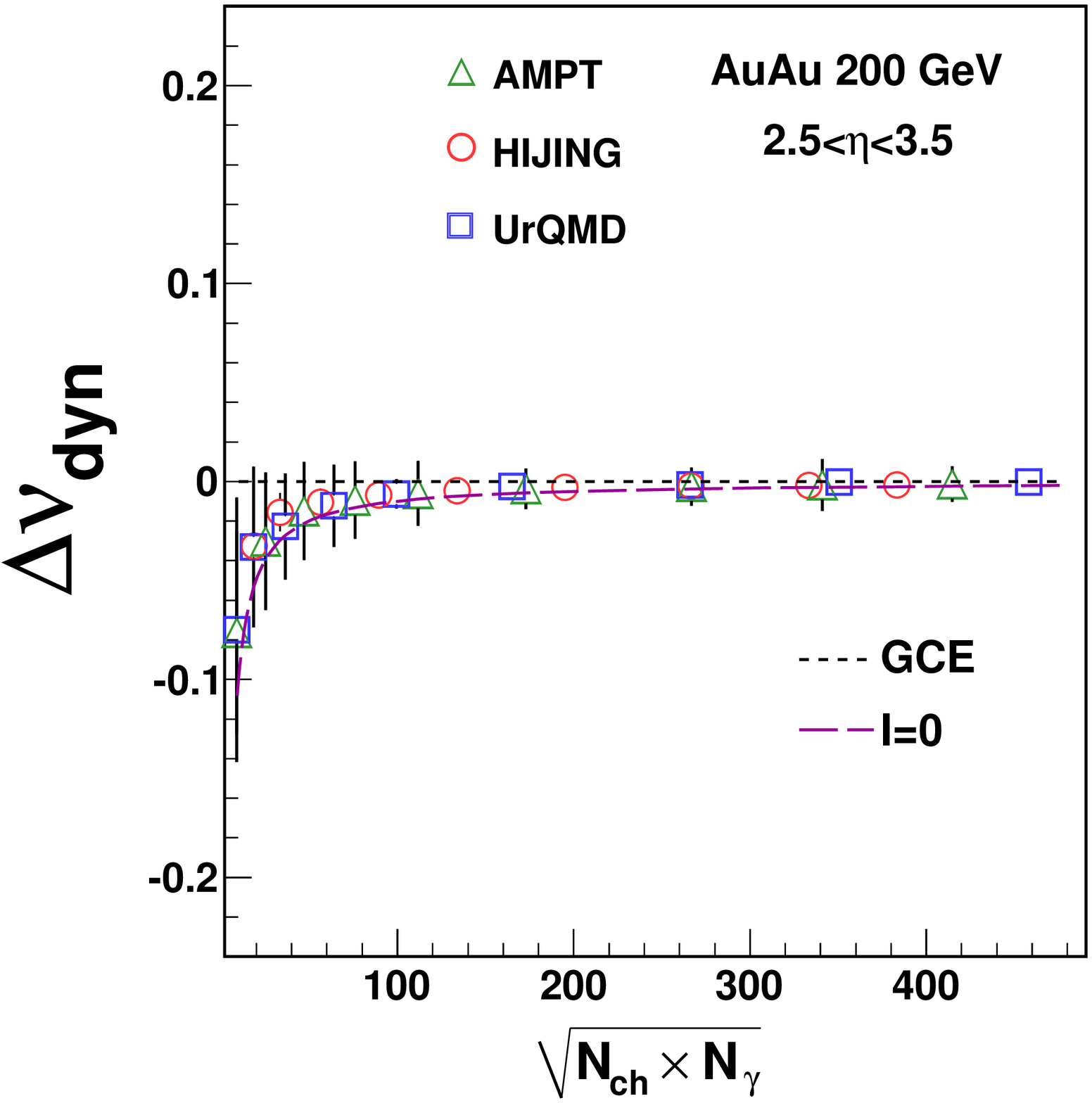}
\label{fig_delndyn_mult}}
\caption{\label{fig_model2} Multiplicity dependence of observables $r_{1,1}$ and $\Delta\ndyn$ as predicted from different models. The curves represent the results from different ensembles of Boltzmann gas of pions from eq.\ref{eq_gce} and eq.\ref{eq_I0} as described in the text. The markers are from different Monte-Carlo models. The error-bars are statistical.}
\end{center}
\end{figure*}
\begin{figure}[h]
\begin{center}
\subfigure[Value of $r_{m,1}$ with $m$ from different models.]{
\includegraphics[height=7cm, width=7cm]{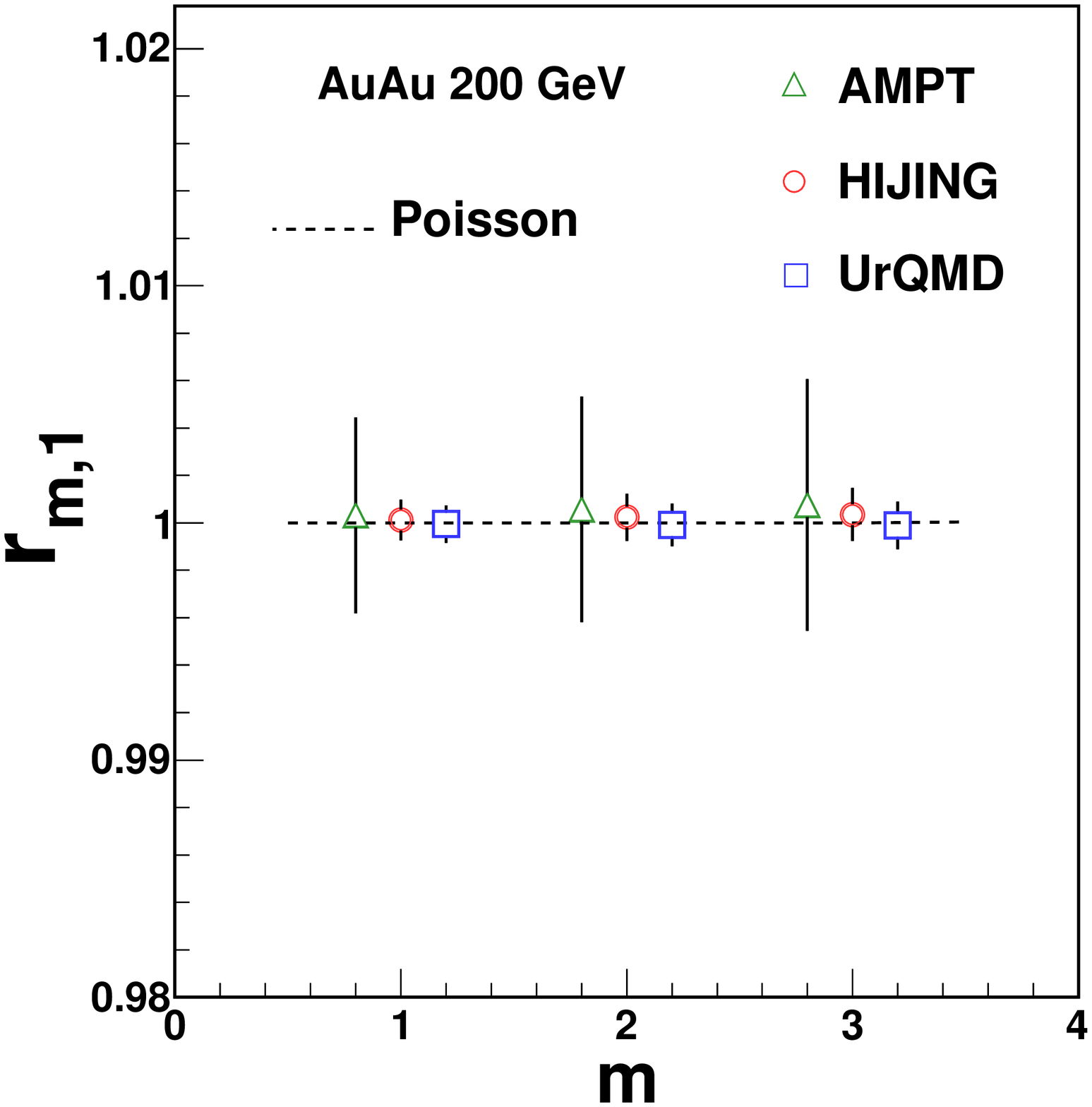}
\label{fig_rm1_model}}
\subfigure[Centrality dependance of variable(markers) $\ndyn$ and $\Delta\ndyn$ and fits(lines) predicted from CLT. Here $N_{\ch}$ and $N_\ph$ refers to the mean multiplicities of charged particles and photons.]{
\includegraphics[height=7cm, width=7cm]{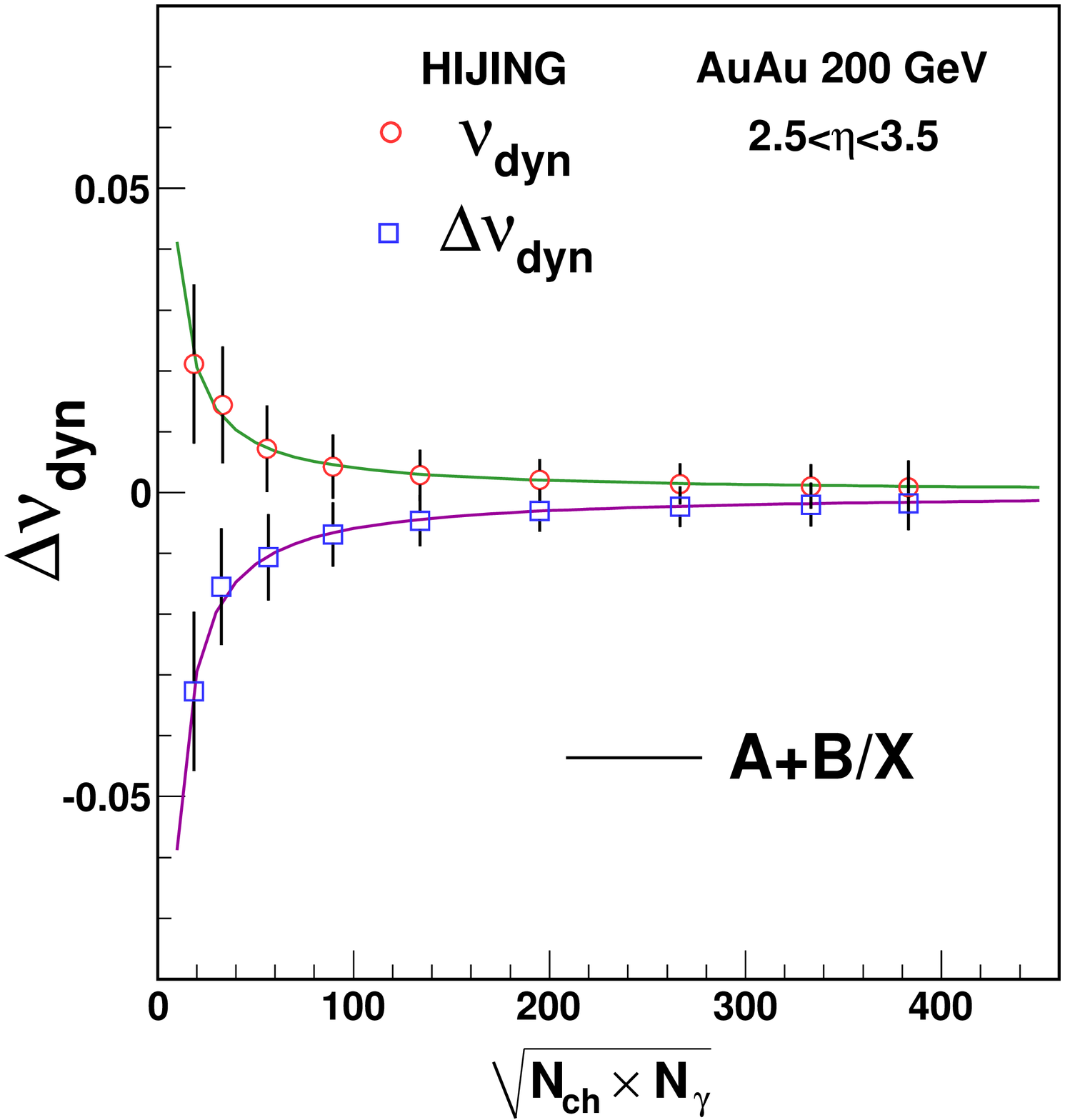}
\label{fig_ndyn_fit}}
\caption{\label{fig_model1}Prediction of variables from different models.} 
\end{center}
\end{figure}
%
\section{Effect of mixture of pion sources}
\label{sec_pifrac}
{\color{Black} In this section we would like to discuss the effect on the observables when event wise pion sources are independent of each other. So far we have considered that in a DCC event,  all the pions detected in a given coverage are coming from the decay of the domains of DCC. 
This assumption might be valid when the detector coverage is same as the size of DCC domains. Let us consider a case when $x$-fraction of events analyzed has DCC like fluctuation carried by $y$-fraction of total pions. This is a realistic scenario when the size of the domain of DCC is smaller than the detector coverage.}
 So for DCC pions we have $\la N \ra_D \,=\, y \la N \ra$ and for generic pions we have $\la N \ra_G \,=\, (1-y) \la N \ra$, $N$ being the total number of pions. The probability to find  $N_D$ pions carrying DCC signal will be given by 
$P(N_{D},N,y)\,=\,^N\!\mathrm{C}_{N_D} \,y^{N_D} (1-y)^{N-N_D}$, which would give $\la N (N-1)\ra_D\,=\, y^2 \la N(N-1)\ra$. 
Now in this case the generating function of eq.\ref{eq_genf} will be replaced by 
\be
G_{obs}\,=\, x^{\prime} G_{DCC} \,+\,x \,G_{\mathrm{DCC}} \, G_{generic}\,+\, (1-x-x^{\prime}) G_{generic}
\ee
in which we view cases with $100\%$ DCC production ($x^\prime$ fraction of events), $100\%$ generic production and a mixture of two as three ``decay modes" of a super cluster. Here $G_\mathrm{DCC}$ has the probability distribution $\pdf=1/2\sqrt{f}$ and $G_{generic}$ has $\pdf= \delta(f-1/3)$. Since we think the case of $100\%$ DCC production is the least realistic, in the following we simplify our expression by taking $x^{\prime}=0$. Now different factorial moments will become functions of $x$ and $y$ {\color{Black}(see appendix-\ref{app_pimix} for detail)}. 
 In this case the observables are modified accordingly , for $\Delta \ndyn$ from eq.\ref{eq_delndyn_simple} will be given by
 \be
 \Delta\ndyn \,=\, \frac{x}{5/9} \,y^2\, \frac{\la N(N-1)\ra}{\la N \ra^2}
 \ee
which consistent with the expression eq.\ref{eq_ndynsig} for $y=1$ case. For Poisson like parent distribution $\Delta\ndyn $ can be expressed as 
\be
 \Delta\ndyn \,=\, \frac{x}{5/9}\,y^2.
\ee
 We note here that $\Delta\ndyn$ still shows the proportionality with the fraction of DCC events $x$. And the interesting fact is that quadratic dependance on $y$ means $\Delta \ndyn$ is more sensitive to the change of fraction of pions carrying DCC-like signals. 

 In similar approach we can express $r_{1,1}$ to be
 \be
  r_{1,1}=\frac{\,5\,-\, 2xy^2}{ 5\,+\,xy^2 }.
 \ee
 This expression is consistent with the approximate expression of $r_{1,1}$ given in Ref.\cite{Mohanty:2005mv} for small values of $x$. The higher order moments will have corrections from higher orders of $y$ which are smaller. To the lowest order approximation, the expression given by eq.\ref{eq_rm1sig} is still valid with fraction $x$ replaced by $xy^2$. 
 \be
r_{m,1}^{\ph-\ch} \approx 1-\frac{\,\,\,m xy^2}{(m+1)} F(m,xy^2)
\label{eq_rm1_pifrac}
\ee
A functional fit of $r_{m,1}$ with $m$ to experimental data by the above expression can restrict the contours of $x$ and $y$.
\section{MODEL PREDICTION}
\label{sec_model}
 In this section we would like to study the behavior of observables from different models available to describe heavy ion data. There are theoretical predictions of isospin fluctuation for a statistical system of pions\cite{ Begun:2004gs, Begun:2010ec}. It can be shown that a system of Boltzmann gas of pions in the grand canonical ensemble (GCE), gives $ \la N_{\pi^0} \ra \, =\,  \la N_{\pi^\pm} \ra ={\mathrm z} $  and one finds mean-square of pion multiplicity and charge-to-neutral pion correlation to be related to mean multiplicities as 
 \bea
 \la N_{\pi^0}^2 \ra \, &=& \, \la N_{\pi^0} \ra + \la N_{\pi^0} \ra^2 \nonumber \\
 \la N_{\pi^\pm}^2 \ra  \, &=& \, \la N_{\pi^\pm} \ra + \la N_{\pi^\pm} \ra^2 \nonumber \\
 \la N_{\pi^0} N_{\pi^\pm} \ra &=&  \la N_{\pi^0} \ra \la N_{\pi^\pm} \ra
 \label{eq_gce}
 \eea
where $\mathrm z$ is the single particle partition function. In ref\cite{Begun:2010ec} it was shown that for an ideal scenario where one assumes the total isospin of the system to be zero, above mentioned relationships will become complicated. An ensemble of the total isospin I=0 gives 
\be
\la N_{\pi^0} \ra \, =\,  \la N_{\pi^\pm} \ra \, =\,  \frac{{\mathrm z}^2}{3}+\frac{{\mathrm z}^3}{6} 
\ee
but the mean-square pions multiplicities are modified as
 \bea
  \la N_{\pi^0}^2 \ra \,&\approx&\, \la N_{\pi^0} \ra + \frac{{\mathrm z}^2}{3}+\frac{{\mathrm z}^4}{15} \nonumber \\
  \la N_{\pi^\pm}^2 \ra \,&\approx&\, \la N_{\pi^\pm} \ra + \frac{{\mathrm z}^4}{10}.
\label{eq_I0}
 \eea
 { \color{Black} We can generalize this result and apply in case of our observables of $\gc$ correlation. The dependance on $\mathrm z$ can be eliminated and final observables can be expressed in terms of experimentally observed quantities like measured multiplicity (say $\sqrt{\la N_\ch \ra \la N_\ph \ra}$).} In this case one has $\la N_\ph \ra \, =\, 2 \la N_{\pi^0} \ra$  and $\la N_\ch \ra \, =\, \la N_{\pi^+} + N_{\pi^-} \ra  \, =\, 2 \la N_{\pi^\pm}\ra.$  Also for decay of neutral pions, in case when all the photons are detected we have\footnote{For Poissonian case $\sigma_\ph= \sqrt{\la N_\ph \ra} = \sqrt{2 \la N_{\pi^0} \ra}=\sqrt{2}  \sigma_{\pi^0}$} $\sigma_{\ph}^2 \, \approx \, 2 \sigma_{\pi^0}^2$.  One can express the mean-square multiplicity to be 
\be
 \la N_\ph^2 \ra =4 \la N_{\pi^0}^2 \ra \,\, ,\,\, 
 \la N_\ch^2\ra  \,= \, 2 \la N_{\pi^\pm}^2  \ra \,+\,  2 \la N_{\pi^+} N_{\pi^-}\ra
 \ee
 and the correlation term will be given by  $\la N_\ph N_\ch \ra \,= \, 4\la N_{\pi^0} N_{\pi^\pm}  \ra$. 
 Now we have
 \bea
 \label{eq_model}
\frac{\mathrm{f}_{20}}{\mathrm{f}_{10}^2}
&=&\frac{1}{2}  \lf  \frac{\npmf}{\la N_{\pi^\pm} \ra ^2}  + \frac{\npm}{\la N_{\pi^\pm} \ra^2} \rf  \nonumber \\
%
\frac{\mathrm{f}_{02}}{\mathrm{f}_{01}^2}
&=&\frac{1}{2}  \lf \frac{\nof}{\la N_{\pi^0} \ra^2}  + 1 \rf
 \nonumber \\
%
\frac{\mathrm{f}_{11}}{\mathrm{f}_{10} \, \mathrm{f}_{01}}
&=&\frac{\nopm}{\la N_{\pi^0} \ra \la N_{\pi^\pm} \ra}
 \eea
 So using  eq.\ref{eq_gce}, eq.\ref{eq_I0} and eq.\ref{eq_model} we can estimate $\ndyn^{\ph-\ch}$ and $r_{1,1}$ for GCE and I=0 systems. Using eq.\ref{eq_delndyn_simple} we can estimate $\Delta\ndyn^{\ph-\ch}$.  For GCE we get from eq.\ref{eq_gce} and eq.\ref{eq_model}, 
 $\ndyn=1/ \sqrt{\la N_\ch \ra \la N_\ph \ra}$, which gives correct multiplicity dependence as predicted from CLT. So we have $\Delta\ndyn^{\ph-\ch}=0$ for GCE. The system of I=0 gives $\Delta\ndyn^{\ph-\ch} \sim -0.98/\sqrt{\la N_\ch \ra \la N_\ph \ra}$ which also agrees with the CLT predictions as shown in fig.\ref{fig_delndyn_mult}. In case of GCE $r_{1,1}$ is predicted to be $2/(1+1/\sqrt{\la N_\ch \ra \la N_\ph \ra})$ which becomes 1 for large values of multiplicity. For system of I=0, $r_{1,1} \sim 1$ for all values of $\sqrt{\la N_\ch \ra \la N_\ph \ra}$ as shown in fig.\ref{fig_r11_mult}.
 
    We have estimated various observables and their centrality dependence using different monte-carlo event generators like HIJING\cite{hijing}, AMPT \cite{ampt} and UrQMD\cite{urqmd} for top RHIC energy. For our calculation we choose one unit of rapidity in forward direction\footnote{both STAR and ALICE experiments has the setup of simultaneous measurements of charged and photon in one unit of rapidity.}. We do the centrality selection based on putting cuts on impact parameter following Glauber model calculation. 
    Fig.\ref{fig_model2} shows the centrality dependence of the observables. The variable $r_{1,1}$ shows flat centrality dependence within error bars. As shown in fig.\ref{fig_r11_mult} and fig.\ref{fig_delndyn_mult}, the results from different monte-carlo models are consistent with each other and the values from the statistical model of Boltzman gas are consistent with other models towards higher multiplicity. At lower multiplicities they have qualitatively different nature probably due to presence of various other effects in the monte-carlo models.

  Fig.\ref{fig_rm1_model} shows the variation of $r_{m,1}$ with its order $m$. Results from all the model are consistent with the generic case of pion production. Fig.\ref{fig_ndyn_fit} shows the centrality dependance of $\ndyn$ and $\Delta\ndyn$ predicted from HIJING.  For comparison of centrality dependance predicted from CLT, we have fitted the points with functional form of $A+B/\sqrt{\la N_\ch \ra \la N_\ph\ra }$. This yields a value of $A\approx5\times10^{-5}$ and $B=-0.6$ for $\Delta\ndyn$. We also note here that the sign of $\Delta \ndyn$ is negative for low multiplicity. This means that the Raw HIJING includes some intrinsic $\ph-\ch$ correlation making the last term of eq.\ref{eq_delndyn} to dominate over individual fluctuation. This can be attributed to the resonance decays present in HIJING model. For DCC like signal sign of $\Delta\ndyn$ should become positive for all centralities.

\section{DCC MODEL}
\label{sec_dccmodel}
 We have tried to implement DCC like anti-correlation signals in HIJING events. For a given event we changed the neutral pion fraction to follow $1/2\sqrt{f}$ like distribution by flipping $\pi^0$ to $\pi^\pm$. And finally we decay the neutral pions to photons. In the process of flipping we make sure that the charge and isospin conservations are maintained. 
\begin{figure}[h]
\includegraphics[height=6cm, width=6cm]{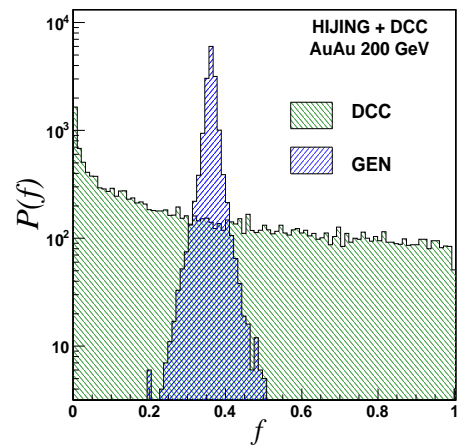}
\caption{\label{fig_dccmodel} Histograms showing distribution of neutral pion fraction for generic and DCC events from HIJING}
\end{figure}
 Fig.\ref{fig_dccmodel} shows the $f$-distribution after the implementation of DCC in HIJING. For generic event the neutral pion fraction is peaked at 1/3 and for DCC events it has a long tail. 
\begin{figure}[h]
\begin{center}
\subfigure[Variation of $r_{1,1}$ with multiplicity]{
\includegraphics[height=7cm, width=7cm]{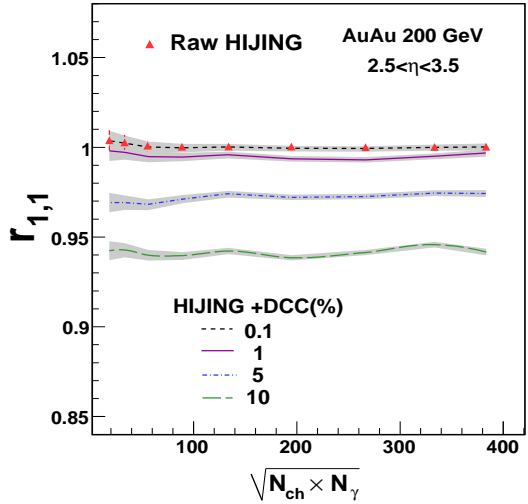}}
\subfigure[Variation of $\ndyn$ with multiplicity]{
\includegraphics[height=7cm, width=7.3cm]{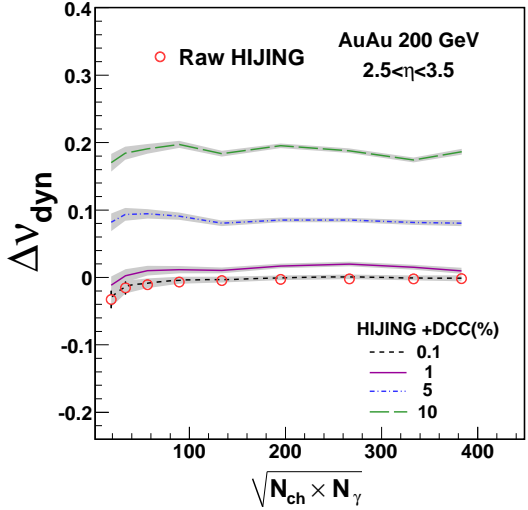}}
\caption{\label{fig_model} Multiplicity dependence of observables $r_{1,1}$ and $\ndyn$ as predicted from DCC implemented HIJING model. Here $N_\ch$ and $N_\ph$ denotes the mean multiplicities of charged particles and photons for various centralities. The gray band shows the statistical error in model calculation.}
\end{center}
\end{figure}
 Since the variation of DCC like domain formation with rapidity and azimuthal angle is not known,  we perform this flipping for all the particles. This produces $1/2\sqrt{f}$ like distribution over all phase space. To make the scenario more realistic we do the calculation of the final variables using total number of detected photons and charged particles rather than considering only pions. Other dominant sources of photons and charged particles include $\eta$, charged kaons and protons respectively. It is difficult to extract the fraction of primordial pions on which the  DCC-like probability distribution could be implemented. HIJING has minijet like environment in   which the production mechanism are ``string fragmentation" and the abundance of particles are weighted by the spin giving large fraction of pions coming from decay of resonances. The primordial pions coming directly from string fragmentation are much smaller.   Alternative environment like hydro models where the massive resonances are exponentially suppressed would give large fraction of soft pions.  The difference between the two models of string fragmentation and hydro is recently contested in ref.\cite{Longacre:2011bx}. We therefore randomly choose pions produced in HIJING events, treat them to be thermal and implement $1/2\sqrt{f}$ distribution.
   %
 \begin{figure}[h]
\begin{center}
\subfigure[Value of $r_{m,1}$ with $m$ for various fraction of DCC events. The solid markers are when $N_\ch$ and $N_\ph$ includes all the charged particles and photons and the hollow markers are when only pions are source of charged particles and photons. The curves are estimations from eq.\ref{eq_rm1sig} and points are from DCC implemented HIJING. ]{
\includegraphics[height=7cm, width=7cm]{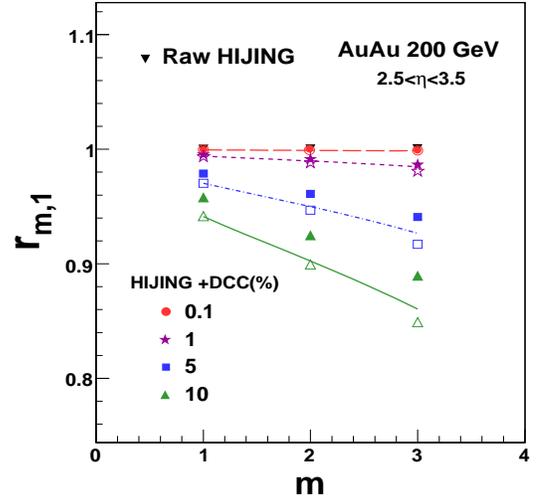}}
\subfigure[Variation of $r_{m,1}$ in DCC events with fraction of pions coming from decay of DCC domains. Curves are estimations from eq.\ref{eq_rm1_pifrac}.]{
\includegraphics[height=7.2cm, width=7cm]{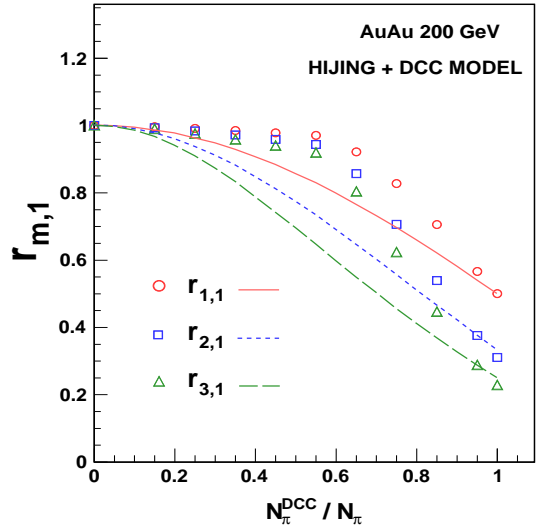}}
\caption{\label{fig_rm1_dccmodel}Sensitivity of $r_{m,1}$ to DCC like signals, estimation shown for $x=1$.}
\end{center}
\end{figure}
 Fig.\ref{fig_model} shows the centrality dependance of the two observables and their sensitivity for different fraction of DCC events. $r_{1,1}$ shows almost flat dependance on multiplicity and we also find similar dependance for all higher moments of $r_{m,1}$.  Absolute values of $r_{1,1}$ are consistent with the prediction ($r_{1,1}=\,(5-2x)/(5+x))$ from eq.\ref{eq_rm1sig}.
 For higher fraction of DCC the centrality dependance has slight non-monotonic behavior. This is also seen in $\Delta\ndyn$. As expected from eq.\ref{eq_ndynsig}, the values of $\Delta\ndyn$ show proportionality with the fraction of DCC events. {\color{Black} The absolute values of $\Delta\ndyn$ are also very close to $\approx x/(5/9)$ as predicted in eq.\ref{eq_ndynsig}. The centrality dependence causes $\approx15\%$ variation of the values of most central to peripheral events for $\Delta\ndyn$.
  Fig.\ref{fig_rm1_dccmodel}(a) shows that the variation of $r_{m,1}$ with $m$.
 The results from model match the theoretical curve (eq.\ref{eq_rm1sig}) when one considers only pions as source 
 of charged particles and photons, however when all other sources are considered the results are off towards lower side. A more detailed study of the sensitivity to fraction of DCC pions is shown in Fig.\ref{fig_rm1_dccmodel}(b). We have shown the sensitivity of $r_{m,1}$ with the fraction of detected pions carrying DCC-signals. 
  In fig.\ref{fig_rm1_dccmodel}(b) we also plot the curves obtained from eq.\ref{eq_rm1_pifrac}. The effect of resonances present in HIJING seems to be resulting in reduced sensitivity of $r_{m,1}$ for lower fraction of DCC pions. 
\section{summary}
\label{sec_sum}
We have developed a procedure for generalization of methods for studying $\gamma$-charge correlation in heavy-ion collisions. One of the primary motivations for this study could be the search for DCC-like anti-correlation signals relevant to the ongoing  heavy ion program at RHIC and LHC. We have discussed the 
robustness of two variables $\Delta\ndyn$ and $r_{m,1}$ and have studied their centrality(multiplicity) dependance. 
The sensitivity of the variables have been studied with the fraction of DCC type events($x$) and the event wise fraction of DCC pions($y$).
These variables where evaluated from different models which do not include the physics of DCC.
We have also developed a Monte-Carlo model where DCC domains have been implemented in HIJING to see the sensitivity of those variables with DCC signals.
 Our results show that the model predictions of the variables are consistent with the theoretical predictions using generating function approach. 
 
 We have implemented the detector effects like efficiency of detection, mis-identification
 to study the effect on observables. The mis-identification factor reduces the effective signal strength for which an approximate expression has been derived in generating function approach. $r_{m,1}$ has been found to be more robust towards mis-identification of photons as compared to $\Delta\ndyn$. The resonance decay can induce correlation which can suppress the anti-correlating DCC signal. A quantitative idea of resonance can be obtained from Monte Carlo model that implements DCC in which we can vary the number of DCC candidates and see the sensitivity of $r_{m,1}$.
 
 We have seen that the variable $\Delta\ndyn$ is highly sensitive to the fractions $x$ and $y$. 
 In a given centrality $\Delta\ndyn$ is proportional to $xy^2$. For generic case of particle production from CLT, it is predicted to be inversely proportional to multiplicity. The sign of $\Delta\ndyn$ would indicate the dominance of correlation over anti-correlation. 

 We also discuss the applicability of the Minimax variable $r_{m,1}$ for heavy-ion collisions. $r_{m,1}$ seem to be flat with centrality. Higher orders of $r_{m,1}$ shows larger sensitivity $x$ and
 can have contribution up to $y^{m+1}$. A simplified form of the functional dependance of $r_{m,1}$ with $m$ has been calculated in generating function approach for lowest order of $y^2$. This would be useful to restrict the signal strength $xy^2$ by fitting the experimental data.
 
\section*{Acknowledgements}
A.T was supported by U.S. Department of Energy under Grants DE- AC02-98CH10886 and DE-FG02- 89ER40531. We thank  R. Longacre, B. Mohanty, C. Pruneau and Y. P. Viyogi for helpful discussions. P.T would like to thank N. R. Sahoo for helping with the MC Models.

\section{APPENDIX}
\subsection{Mixed events}
\label{app_mixevnt}
While analyzing data sample to calculate $\ndyn^{\ph-\ch}$, one can estimate the generic term by doing a mixed event analysis. A simple method we prescribe is to take total number of photons and total number of charge particles from different events would only effect correlation terms like $\mathrm f_{11}$ in $\ndyn^{\ph-\ch}$ keeping other factorial moments unchanged. In such case we must have 
\be
\left. \frac{\mathrm{f}_{11}}{\mathrm{f}_{10} \, \mathrm{f}_{01}} \right|_{mixed} \, \approx \, \frac{\la N(N-1)\ra}{\la N \ra^2}.
\ee
Taking a particular combination of factorial moments we can calculate the generic value for $\ndyn$ we need to calculate $\Delta\ndyn$ 
For example one can show that 
\be
\left . \lf 3 \frac{\mathrm{f}_{11}}{\mathrm{f}_{10} \, \mathrm{f}_{01}}
\,-\, 4 \frac{\mathrm{f}_{20}}{\mathrm{f}_{10}^2}\,+\,\frac{\mathrm{f}_{02}}{\mathrm{f}_{01}^2} \rf \right |_{mixed} \,=\, \frac{1}{2\la f \ra \la N \ra}
\ee
which is equal to $\ndyn^{generic}$. But in case of contamination effects present in the data sample one cannot apply this simple method since in that case the efficiency terms cannot be eliminated from $\ndyn$. A full GEANT simulation with a known event generator which doesn't include the physics of DCC is suggested to estimate the generic value of $\ndyn$.

\subsection{Mis-identification}
\label{app_misid}
In case of mis-identification of photon as charge particles and vice-versa the fractorial moments are modified as given in eq.\ref{eq_fmom_eff}. 
The observables $\Delta\ndyn$ and $r_{m,1}$ will be given by
%
\begin{widetext}
\centering
\bea
&&\Delta\ndyn^{\ph-\ch}=\nonumber \\
&&\ngap\! \lf \! \frac{\la \lf (1-f)\eff_\ch+2f {\eff}_{\ph,\ch} \rf^2 \ra}{\la (1-f)\eff_\ch \!+\! 2f {\eff}_{\ph,\ch} \ra^2}
\!+\!\frac{ \la \lf (1-f)\eff_{\ch,\ph} +2f \eff_\ph \rf^2 \ra}{\la (1-f)\eff_{\ch,\ph} +2f \eff_\ph \ra^2}
\!-\!2 \frac{\la \lf (1-f)\eff_{\ch} \! + \! 2f  {\eff}_{\ph,\ch} \rf  \lf(1-f)\eff_{\ch,\ph}\!+\!2f\eff_\ph \rf \ra}{\la (1-f)\eff_{\ch} \! + \! 2f {\eff}_{\ph,\ch} \ra  \la(1-f)\eff_{\ch,\ph}\!+\!2f\eff_\ph \ra } \! \rf \! \frac{\la N(N-1)\ra}{\la N \ra^2} \nonumber \\
%
%
%
\eea
\bea
&&r_{m,1} \,=\, \frac{\la N(N\,-\,1) \, \lf (1\,-\, f) \eff_{\ch}\,+\,2f {\eff}_{\ph,\ch} \rf^m  \, \lf(1\,-\,f)\eff_{\ch,\ph}\,+\,2f\eff_\ph \rf \,+\, 2Nf \eff_\ph {\eff}_{\ph,\ch}  \ra \, \la(1\,-\,f)\eff_\ch\,+\,2f {\eff}_{\ph,\ch} \ra}{\la N(N-1) \, \lf (1-f)\eff_\ch+2f {\eff}_{\ph,\ch} \rf ^{m+1} \, + \, 2Nf \, \lf 2\eff_{\ph,2\ch} +{\eff}_{\ph,\ch}^2\rf \ra \la (1-f)\eff_{\ch,\ph} \, + \, 2f\eff_\ph \ra} \nonumber \\
%
%
\eea
\end{widetext}
 In case of $\eff_{\ph,\ch}=\eff_{\ph,2\ch}=0$ one recovers  eq.\ref{eq_delndyn_eff} and eq.\ref{eq_rm1_eff}.
\subsection{Pion mixture}
\label{app_pimix}
In case of $x$-fraction of DCC events containing $y$-fractions of pions carrying DCC signal,  different factorial moments are given by
\bea
\mathrm{f}_{10}&=&\la 1-f \ra \eff_\ch \na \nonumber \\
%
\mathrm{f}_{01}&=&\la f \ra 2 \eff_{\ph} \la N \ra \nonumber
\eea
which is same as the case corresponding to $y=1$.  But higher order moments are modified to be
\begin{widetext}
\bea
\mathrm{f}_{11}&=& \lf 2x y (1-y) \la f \ra \la 1-f  \ra \la N \ra^2  \,+  \, ( (1-xy(2-y)) \la f(1-f)\ra_G  \,+\, xy^2\la f(1-f)\ra_D )\la N\lf N-1\rf \ra \rf 2\eff_{\ph} \, \eff_\ch \nonumber \\
%
\mathrm{f}_{20}&=& \lf 2xy(1-y)  \la 1-f \ra^2  \la N \ra^2 \, + \, ( (1-xy(2-y)) \la (1-f)^2 \ra_G  \,+\, xy^2\la (1-f)^2 \ra_D ) \la N \lf N-1 \rf \ra \rf  \eff_\ch^2 \nonumber \\
\mathrm{f}_{02}&=& \lf 2xy(1-y)  \la f \ra^2  \la N \ra^2 \, + \, ( (1-xy(2-y)) \la f^2 \ra_G  \,+\, xy^2\la f^2 \ra_D ) \la N \lf N-1 \rf \ra \rf  4 \eff_{\ph} ^2  +  2\eff_{\ph}^2 \la f \ra \la N \ra  \nonumber \\
\eea
\end{widetext}
which gives 
\bea
  r_{1,1}&=&\frac{\,5-\, 2xy^2}{ \,5\,+\,xy^2 } \nonumber \\
  r_{2,1}&=&\frac{\,35\,-\, xy^2(21-4y)}{ 35\,+\,xy^2(21-2y) }
\eea
and so on. The general formula for $r_{m,1}$ is given by
\be
r_{m,1}\,=\, 1-\frac{\,\,\,m xy^2}{(m+1)} F(m,xy^2)+{{\cal O}(xy^3)\, \cdots}
\ee
in which $r_{m,1}$ will have contribution up to $xy^{m+1}$. Since $y\le1$ higher order contribution of $y$ are smaller and the approximate form of the above expression would be given by
\be
r_{m,1}\,\approx \, 1-\frac{\,\,\,m xy^2}{(m+1)} F(m,xy^2)
\ee
where $F(m,xy^2)$ is given by eq.\ref{eq_fmx}.


\begin{thebibliography}{50}

\bibitem{Jeon:1999gr}
  S.~Jeon and V.~Koch,
  Phys.\ Rev.\ Lett.\  {\bf 83}, 5435 (1999)
  [arXiv:nucl-th/9906074].

\bibitem{Bjd} J.D. Bjorken, What lies ahead?, SLAC-PUB-5673, 1991.


\bibitem{Blaizot:1992at}
  J.~P.~Blaizot and A.~Krzywicki,
  Phys.\ Rev.\  D {\bf 46}, 246 (1992).

\bibitem{Rajagopal:1992qz}
  K.~Rajagopal and F.~Wilczek,
  Nucl.\ Phys.\  B {\bf 399}, 395 (1993)
  [arXiv:hep-ph/9210253].

\bibitem{Rajagopal:1995bc}
 K.~Rajagopal,
 arXiv:hep-ph/9504310.

\bibitem{Brooks:1999xy}
  T.~C.~Brooks {\it et al.}  [MiniMax Collaboration],
  Phys.\ Rev.\  D {\bf 61}, 032003 (2000)
  [arXiv:hep-ex/9906026].

\bibitem{Lattes:1980wk}
  C.~M.~G.~Lattes, Y.~Fujimoto and S.~Hasegawa,
  Phys.\ Rept.\  {\bf 65}, 151 (1980).

\bibitem{Aggarwal:1997hd}
  M.~M.~Aggarwal {\it et al.}  [WA98 Collaboration],
  Phys.\ Lett.\  B {\bf 420}, 169 (1998)
  [arXiv:hep-ex/9710015].

\bibitem{Aggarwal:2000aw}
  M.~M.~Aggarwal {\it et al.}  [WA98 Collaboration],
  Phys.\ Rev.\  C {\bf 64}, 011901 (2001)
  [arXiv:nucl-ex/0012004].
  
\bibitem{Aggarwal:2002tf}
  M.~M.~Aggarwal {\it et al.}  [WA98 Collaboration],
  Phys.\ Rev.\  C {\bf 67}, 044901 (2003)
  [arXiv:nucl-ex/0206017].

\bibitem{Collaboration:2011rsa}
  M.~M.~Aggarwal {\it et al.},
  Phys.\ Lett.\  B {\bf 701}, 300 (2011)
  [arXiv:1103.2489 [nucl-ex]].

\bibitem{Appelshauser:1999ft}
  H.~Appelshauser {\it et al.}  [NA49 Collaboration],
  Phys.\ Lett.\  B {\bf 459}, 679 (1999)
  [arXiv:hep-ex/9904014].

\bibitem{Krzywicki:1998sc}
  A.~Krzywicki and J.~Serreau,
  Phys.\ Lett.\  B {\bf 448}, 257 (1999)
  [arXiv:hep-ph/9811346].


\bibitem{Rajagopal:2000yt}
  K.~Rajagopal,
  Nucl.\ Phys.\  A {\bf 680}, 211 (2000)
  [arXiv:hep-ph/0005101].

\bibitem{STARNIM:PMD}
M.M.~Aggarwal {\it et al.} 
Nucl.\ Instrum.\ Meth. A {\bf 499}, 751 (2003)
%

\bibitem{STARNIM:FTPC}
K.H.~Ackermann {\it et al.} 
Nucl.\ Instrum.\ Meth. A {\bf 499}, 713 (2003)
%

\bibitem{Aamodt:2008zz}
  K.~Aamodt {\it et al.}  [ALICE Collaboration],
  JINST {\bf 3}, S08002 (2008).



  
\bibitem{nudyn}
  C.~Pruneau, S.~Gavin and S.~Voloshin,
  Phys.\ Rev.\  C {\bf 66}, 044904 (2002)
  [arXiv:nucl-ex/0204011].

\bibitem{star_kpi}
  B.~I.~Abelev {\it et al.}  [STAR Collaboration],
  Phys.\ Rev.\ Lett.\  {\bf 103}, 092301 (2009)
  [arXiv:0901.1795 [nucl-ex]].

\bibitem{star_dcc}
  S.~M.~Dogra  [STAR Collaboration],
  J.\ Phys.\ G {\bf 35}, 104094 (2008).

\bibitem{Minimax}
  T.~C.~Brooks {\it et al.}  [MiniMax Collaboration],
  Phys.\ Rev.\  D {\bf 55}, 5667 (1997)
  [arXiv:hep-ph/9609375].

\bibitem{Pumplin}
  J.~Pumplin,
  Phys.\ Rev.\  D {\bf 50}, 6811 (1994)
  [arXiv:hep-ph/9407332].

\bibitem{Luo:2010by}
  X.~F.~Luo, B.~Mohanty, H.~G.~Ritter and N.~Xu,
  J.\ Phys.\ G {\bf 37}, 094061 (2010)
  [arXiv:1001.2847 [nucl-ex]].
  
\bibitem{Mohanty:2005mv}
 B.~Mohanty and J.~Serreau,
 Phys.\ Rept.\  {\bf 414}, 263 (2005)
  [arXiv:hep-ph/0504154].
  
\bibitem{Begun:2004gs}
  V.~V.~Begun, M.~Gazdzicki, M.~I.~Gorenstein and O.~S.~Zozulya,
  Phys.\ Rev.\  C {\bf 70}, 034901 (2004)
  [arXiv:nucl-th/0404056].
\bibitem{Begun:2010ec}
  V.~V.~Begun, M.~I.~Gorenstein and O.~A.~Mogilevsky,
  Phys.\ Rev.\  C {\bf 82}, 024904 (2010)
  [arXiv:1004.2918 [nucl-th]].


\bibitem{hijing}
  X.~N.~Wang and M.~Gyulassy,
  Phys.\ Rev.\ D {\bf 44}, 3501(1991).
  
 \bibitem{ampt}
  Z.~W.~Lin, C.~M.~Ko, B.~A.~Li, B.~Zhang and S.~Pal,
  Phys.\ Rev.\  C {\bf 72}, 064901 (2005)
  [arXiv:nucl-th/0411110].
  
\bibitem{urqmd}
  S.~A.~Bass {\it et al.},
  Prog.\ Part.\ Nucl.\ Phys.\  {\bf 41}, 255 (1998)
  [Prog.\ Part.\ Nucl.\ Phys.\  {\bf 41}, 225 (1998)]
  [arXiv:nucl-th/9803035].
  
      
\bibitem{Longacre:2011bx}
  R.~S.~Longacre,
  arXiv:1105.5321 [nucl-th].




\end{thebibliography}
\end{document}